\documentclass[review]{elsarticle}
\usepackage[legalpaper, margin=1in]{geometry}
\usepackage{graphicx}
\usepackage{amssymb}
\usepackage{lineno}
\usepackage{lineno,hyperref}
\usepackage{amsmath}
\modulolinenumbers[5]
\usepackage{subfig}
\usepackage[figurename=Fig.]{caption}
\usepackage{booktabs,tabularx}

\journal{Ocean Engineering}

\begin{document}
\begin{frontmatter}


\title{The hydrodynamic characteristics of Autonomous Underwater Vehicles in rotating flow fields}



\author{A. Mitra\fnref{myfootnote}} \author{J.P. Panda} \author{H.V. Warrior}
\address{Department of Ocean Engineering and Naval Architecture,
IIT Kharagpur, India}

\fntext[myfootnote]{corresponding author: arindam.mitra@iitkgp.ac.in }

\begin{abstract}
In this article, the hydrodynamic characteristics of Autonomous Underwater Vehicles (AUVs) are investigated and analyzed under the influence of rotating flow fields, that were generated in a recirculating water tank via a rotating propeller. Initially, experiments were carried out for the measurement of flow field variables and Quantities of Interest across the AUV in the presence of the rotating propeller while varying the rotational speed and the extent of rotational flow strength. The flow field across the AUV was measured using an Acoustic Doppler Velocimeter (ADV). These measured turbulent flow statistics were used to validate the Reynolds Stress Model (RSM) based numerical predictions in a commercial CFD solver. After preliminary validation of the turbulent flow statistics with the numerical predictions, a series of numerical simulations were performed to investigate the effect of the rotational flow field of the propeller on the drag, skin friction and pressure coefficients of the AUV. The operating speed and location of the propeller were also varied to check their affects on the hydrodynamic performance of AUV. The results provided in this article will be useful for the design optimization of AUVs cruising in shallow water where the flow is highly rotational because of wave-current interactions. Additionally the results and analysis are relevant to study the design and operation of AUVs that have to operate in a group of unmanned underwater vehicles or near submarines and ships where the flow field is highly complex and such rotational effects are present.     
\end{abstract}

\end{frontmatter}


\section{Introduction}
\label{S:1}
Autonomous underwater vehicles has significant applications in underwater activities such as mining, geophysical surveys, deep-sea exploration of hydrocarbons, maintenance of underwater pipelines and floor planning activities related to installation of underwater structures. Most of these activities are usually carried out in deeper oceans \citep{panda2020review,SAHOO2019145}, where the flow field is complex and rotational. The oceanic flow field is also highly complex and rotational near submarines, ships and large remotely operated vehicles \citep{leong2015quasi,leong2015numerical} and near free surface \citep{Tian2019,amiri2018does} because of wave-current interactions. 

Accurate prediction of hydrodynamic parameters of the AUVs is closely related to their safety, stability and control in the regions where complex flow field exists\citep{jagadeesh2009experimental}. Proper understanding of detailed evolution of the hydrodynamic parameters in complex operating conditions will lead to improved design of propulsion, control and navigation system of the AUVs \citep{SAHOO2019145}. Fluid dynamic experiments in deeper oceans are not a cost effective approach for the testing of hydrodynamic parameters. However with increase in computational facilities in recent years, computational fluid dynamics (CFD) based numerical simulations can be considered as an alternate approach for the fluid dynamic studies related to AUVs \citep{huang2019effects,jagadeesh2005application,mansoorzadeh2014investigation}.

There are few standard experimental studies available in literature, in which hydrodynamic studies are performed for autonomous underwater vehicles\citep{SAHOO2019145}. \cite{jagadeesh2009experimental} have performed a series of benchmark experiments on AUVs in which, the hydrodynamic parameters such as drag, lift and pitching moment coefficients of an AUV were analyzed in a towing tank. In these experimental studies it was noted that the drag, lift and pitching moment coefficients are highly responsive to operating speed and angles of attack. \cite{mansoorzadeh2014investigation} have studied the effect of the free surface on the hydrodynamic parameters of the AUV. The variation of the coefficients of drag and lift of the AUV was studied at different submergence depths. In this study it was observed that the total drag of the AUV is higher near the free surface because of the larger wave making resistance. \cite{saeidinezhad2015experimental} performed experimental studies on the behaviour of an underwater vehicle model with a non-axisymmetric nose in pitch maneuver. The model was fixed in a  wind tunnel at high Reynolds Numbers. \cite{mitra2019effects} have analyzed the the effects of turbulence on the drag and lift coefficients of the AUV. The flow field across the AUV was measured and used to validate CFD based numerical simulations. It was observed that the free stream turbulence reduces the drag of the AUV by suppressing its flow re-circulation zone. The results of drag and lift coefficients were also provided for different angles of attack. \cite{amiri2019urans} have studied the hydrodynamic characteristics of shallowly submerged underwater vehicle at steady drift. \cite{amiri2020initial} evaluated the effect of free surface on the maneuverability of underwater vehicles.

There is an inherent limitation to the efficacy, applicability and utility of such experimental investigations of AUV performance under varied and realistic operating conditions. The experimental procedures are time consuming and are not cost effective. For many cases the operating conditions cannot be replicated to a high fidelity without incurring high costs. Most of the interesting hydrodynamic studies on the AUVs are performed using computational fluid dynamics based techniques by using different turbulence models \citep{bao2019resistance}. Such turbulence models reduce the computational costs and the wall time for such simulations while still delivering results that are useful for engineering design. The basic computational fluid dynamics treatment of turbulence can be mainly classified as eddy viscosity models\citep{warrior2014b,maity2011reynolds,Panda2021asimple,panda2020review1} (mainly the two equation models)\citep{kimura2003non}, Reynolds stress models (popularly known as Reynolds stress transport models)\citep{panda2018experimental,mishra2010dynamical,warrior2014}, large eddy simulation and direct numerical simulations. \cite{jagadeesh2005application} used different low Reynolds number two equation turbulence models to predict the flow characteristics along the AUVs. \cite{tyagi2006calculation} used eddy viscosity based two equation turbulence models to calculate the moment and transverse hydrodynamic 
damping force coefficients of a typical AUV model. \cite{salari2017numerical} numerically studied flow hydrodynamics of an AUV moving closer to the free surface by using different class of turbulence models, one is the two equation $k-\epsilon$ model and another is the Menter $k-\omega$ SST model, which transitions between different models based on the flow conditions. \cite{chen2017computational} studied stability of an underwater helicopter (a different class of AUV) by using RANS based simulations. \cite{mostafapour2018effects} have studied the effect of Reynolds number on the hydrodynamic characteristics of AUV at different Reynolds numbers using a two equation turbulence model. \cite{Geng2018} have studied the hydrodynamic characteristics of a synthetic jet steered AUV numerically in lateral and yaw motion. \cite{mitra2020experimental} studied the hydrodynamic characteristics of an AUV hull over sea-beds with complex terrains. Mainly the bed slope of the channel was varied to test its effect on the flow field of the AUV and subsequently on its hydrodynamic parameters. It was noticed that, with increase in bed slope the drag of the AUV increases. \cite{da2017numerical} analyzed drag, lift and torque coefficients of an AUV using large eddy simulations(LES) at different angles of attack. \cite{dasilva2019} have performed numerical simulations for an AUV submitted ocean currents. More recently, \cite{zhang2019investigation}investigated the hydrodynamic characteristics of an multi AUV system operating in tandem. Mainly the Resistance characteristics was studied using SST turbulence model. \cite{Tian2019} studied the effect of free surface on the drag coefficient of the AUV in coupled wave current flows. The wave effects were analyzed for different Reynolds numbers, wave heights and submergence depths. It was noticed that the lift varied significantly in presence of wave and the drag of AUV increases with increase in wave height. More recently \cite{panda2021numerical} have studied the drag reduction of an underwater vehicle model using anti-turbulence surfaces.

Although various numerical and experimental studies are available in the literature in which the hydrodynamic parameters are studied under varying operating conditions like near submarines, near free surface, near sea bottom, near sea-beds with complex topography and in presence of free stream turbulence, there is no such result available in which the hydrodynamic parameters of the AUV is studied in a rotating flow field. Rotationally dominated flow fields are fundamentally different from other varities of flow fields with respect to the manner and the extent by which they may affect AUVs. As an example, wingtip vortices in the turbulent wake of aircraft pose grave hazards to other aircraft especially during their take-off and landing phases \citep{crouch2001active}. If there has not elapsed an adequate period to allow this wingtip vortex to decay, the subsequent aircraft can be destabilized by the wingtip vortex, leading to grave consequences. Wake turbulence warnings are issued at airports to ensure adequate separation between aircraft to ensure enough time for the decay of these vortices \citep{rossow1999lift}. In this article, the hydrodynamic parameters such as, drag, skin friction and pressure coefficient variations are studied by placing the AUV in a rotating flow field. The rotating flow field in the vicinity of AUV was produced by using a propeller. The propeller speed was varied by using single phase DC motor controller. An acoustic doppler velocimeter (ADV) was used to calculate the turbulence statistics near the AUV hull. Those turbulent flow statistics such as turbulence kinetic energy and turbulent shear stress (shear stresses has direct effect on the evolution of wall parameters) profiles were used to validate the Reynolds stress model prediction of flow field along the AUV. The Reynolds stress models were used instead of two equation models, since the former has highest potential to replicate the non-local flow dynamics resulting from stream line curvature and can capture the rotational effects more accurately\citep{panda2020review}. After these preliminary validations series of numerical simulations were performed by varying the rotational speed of the propeller to generate different rotational flow fields of varying strengths. The effect of the rotating flow field of varying strength on the hydrodynamic parameters of the AUV was studied using numerical simulations. The distance of the propeller from the AUV was also varied to check the effect of distance on the evolution of hydrodynamic parameters along the AUV.       

\section{Experimental setup}
\label{S:2}
\begin{figure}
\centering
\subfloat[Schematic of the recirculating water tank (all dimensions are in Millimeter)]{\includegraphics[height=5cm]{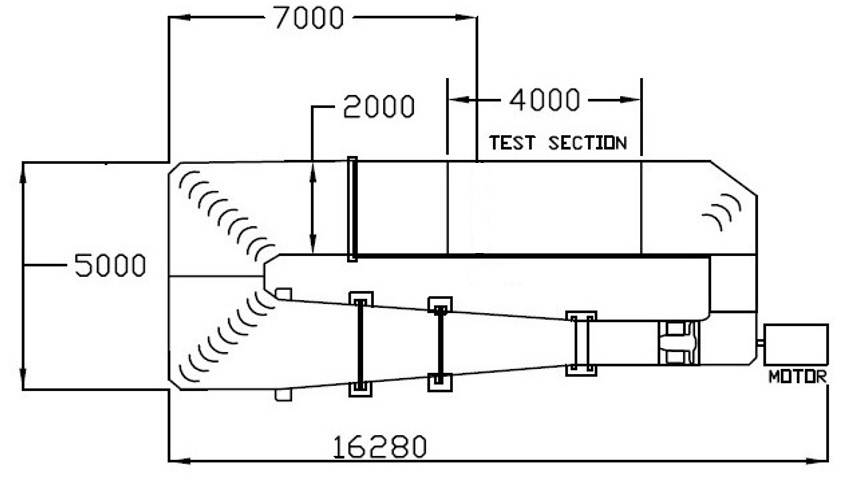}}\\
\subfloat[Experimental setup in the recirculating water tank, The detailed arrangement of AUV hull, propeller and the ADV.]{\includegraphics[height=6cm]{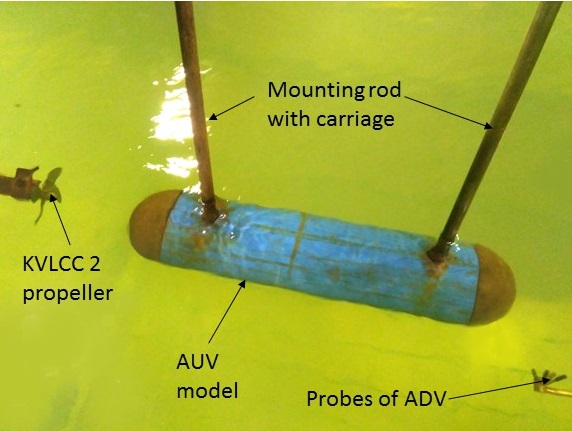}}\\
\subfloat[Schematic of the detailed arrangement of AUV hull and propeller (size not to scale). The arrow marks represent flow direction. ]{\includegraphics[height=10cm]{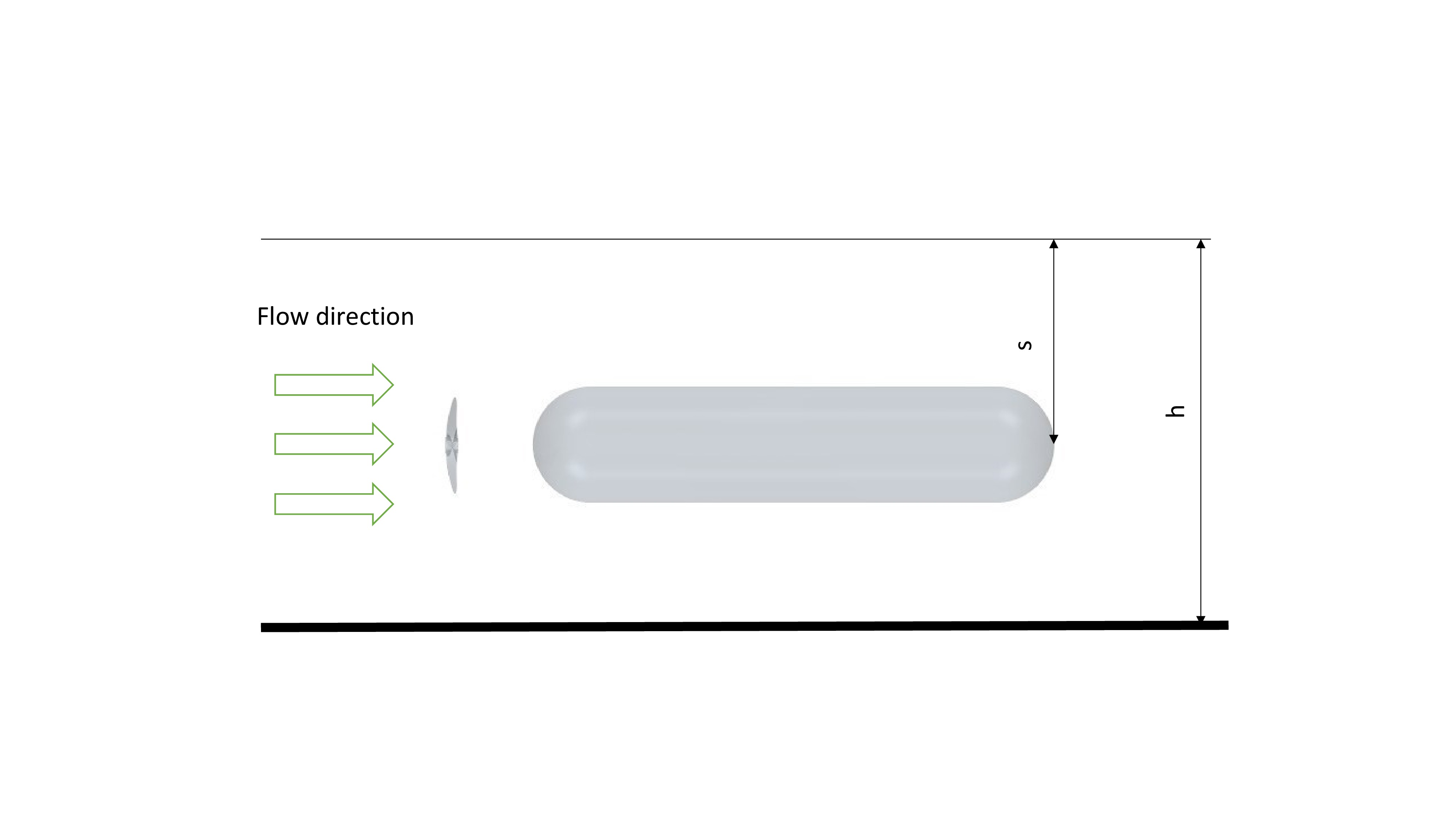}}\\
\caption{Experimental setup in the recirculating water tank. \label{fig:1}}
\end{figure}

The experiments were performed at a recirculating open water tank at the Indian Institute of Technology, Kharagpur. The length of the test section of the channel is $6$ meters. The width and depth of the water tank are 2 and 1.5 meters respectively. For all cases of our experiments the depth of water was taken as 1 meter. In all the experiments and numerical simulations x, y and z correspond to the main flow, transverse and vertical directions respectively. The schematic diagram of the tank is shown in  figure \ref{fig:1}a. The detailed arrangement of the AUV and propeller is shown in figure\ref{fig:1}b. The AUV hull was placed in the test section of the recirculating water tank. The diameter and length of the AUV hull was $0.1$ and $0.5$ meter respectively. The AUV has a cylindrical shape with two hemispherical ends and its shape is similar to the configuration utilized in the investigation of \citep{mansoorzadeh2014investigation}. Although several complicated shapes are available in literature\citep{dasilva2019}, we have considered a simpler shape, since our primary focus was to study the effect the external flow field on its hydrodynamic performance. The AUV was fixed at a depth of 0.3 meter from the free surface. The distance of AUV from the propeller was varied from 0.08m to 0.23m, to check its effect on the evolution of hydrodynamic parameters of the AUV. 

The KVLCC2 propeller(1:100 scale) was used to generate rotating flow field in the vicinity of the AUV hull. The diameter of the propeller was $0.098$ meters. A Single phase DC motor of 0.5 HP was used to rotate the propeller. The RPM of the motor was controlled by a control panel, where mainly voltage range was regulated. The propeller was rotated at 3 different RPMs (800, 1000 and 1200 respectively). A tachometer was used to measure the RPM of the shaft connecting the motor with propeller. Schematic of the detailed arrangement of AUV hull and propeller are shown in figure \ref{fig:1}c. The arrow marks in the diagram represent the flow direction. 

\begin{figure}
\centering
\includegraphics[height=4cm]{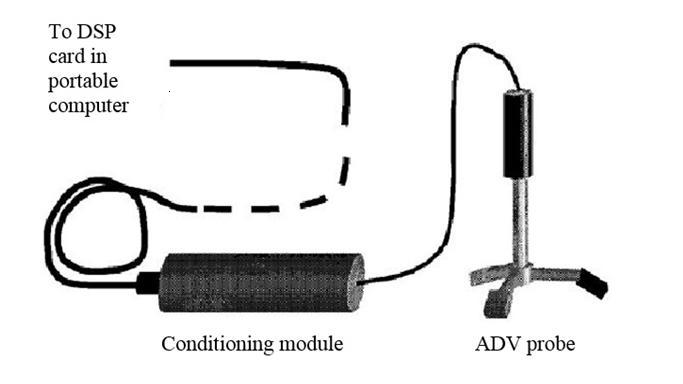}
\caption{Schematic of the acoustic doppler velocimeter} \label{fig:2}
\end{figure}

An acoustic doppler velocimeter (ADV) was used in our experiments to measure the instantaneous flow velocities. The ADV works on the principle of Doppler's shift. A schematic of ADV is shown in figure \ref{fig:2}. The ADV system always suffer from errors due to probe orientation, sampling frequency, instrument velocity range and local flow properties. The major sources of errors are sampling error, error due to noise and velocity gradients in the sampling volume. The detailed methodology of finding uncertainties in ADV system is available in \cite{voulgaris1998evaluation}. The measured mean flow velocities have errors within $1\%$ of the measured value\citep{voulgaris1998evaluation}. For our experiments a sampling rate of 200 Hz was considered and at each location the ADV was fixed for 6 minutes to collect velocity statistics.

A series experiments were conducted with all these experimental set-ups by varying the rotational speed of propeller. The flow velocity was maintained at 0.42 m/s for all the cases of experiments. The ADV was fixed over a frame and allowed to move over a roller arrangement to collect velocity statistics at six different locations across the AUV. All those points are at a distance of 0.05 meter from the AUV wall. From the velocity statistics, we have calculated the mean and fluctuating velocity for the three different directions and obtained the turbulence parameters such as turbulence kinetic energy and 6 independent components of the Reynolds stresses. We only have utilized the turbulence kinetic energy and Reynolds shear stresses for the validation of the numerical model.    

\begin{figure}
\centering
\subfloat[The computational domain]{\includegraphics[height=8cm]{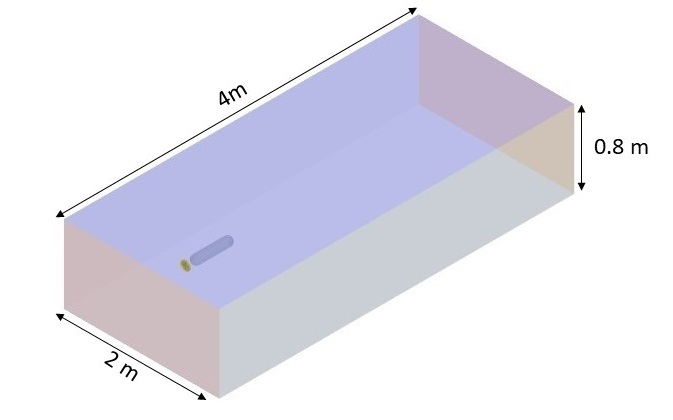}}\\
\subfloat[The mesh]{\includegraphics[height=8cm]{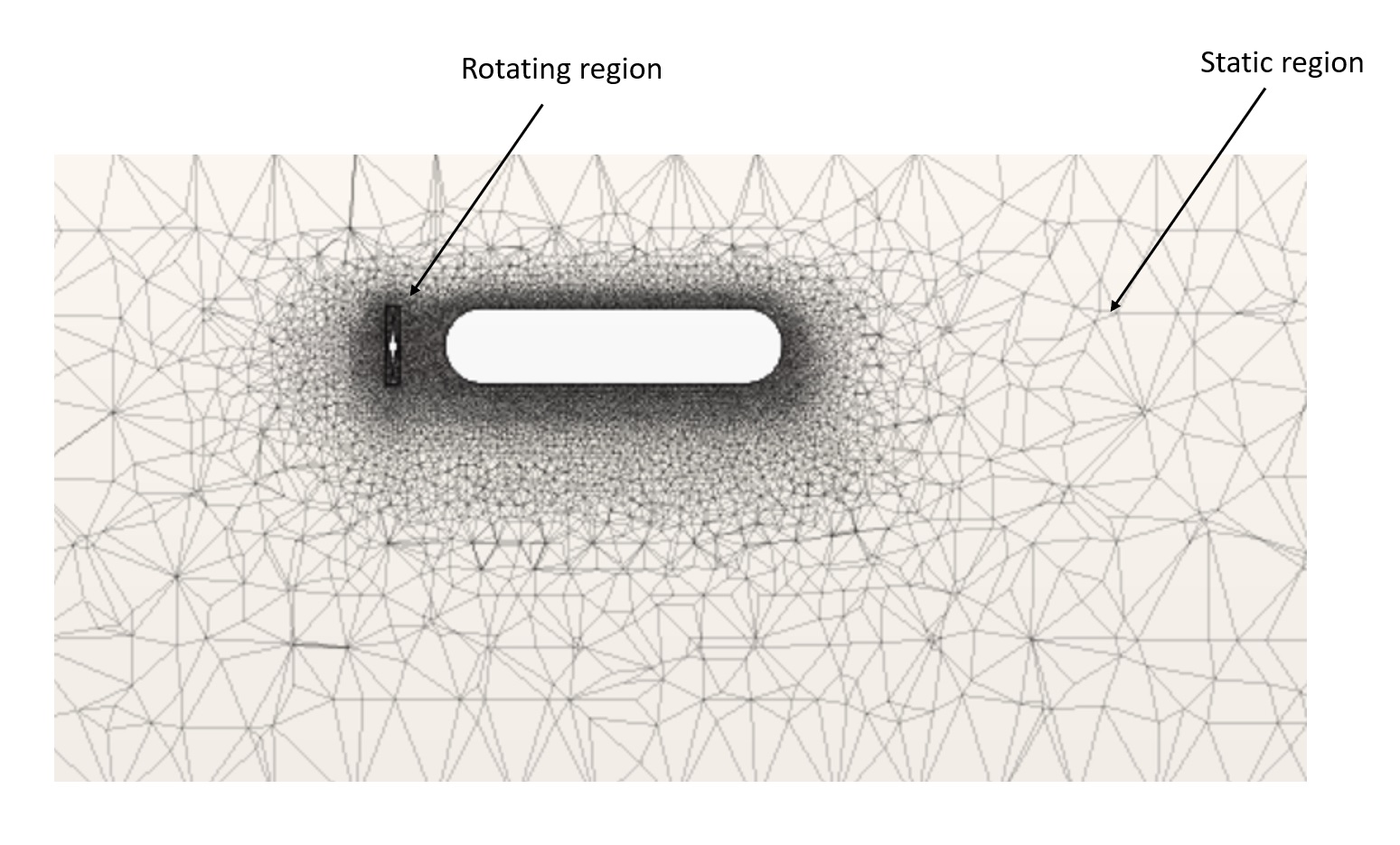}}\\
\subfloat[The mesh near the propeller]{\includegraphics[height=8cm]{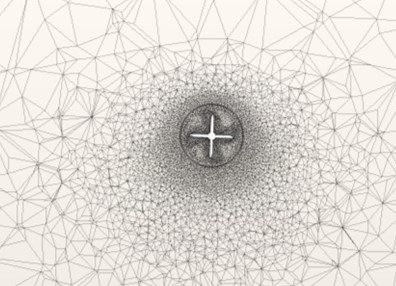}}\\
\caption{Schematic of the computational domain and pictures of mesh near the AUV and propellor.\label{fig:2k}}
\end{figure}
\section{Numerical modelling details}
\label{S:3}
\subsection{Fluid domain and mesh}
The domain is a rectangular computational space, which is presented in Fig. 3a.  Its size is $4m \times 2m \times 0.8m$. The domain length is approximately five times the AUV length behind the stern of the AUV and two times the AUV length in front of AUV. In fig. \ref{fig:2k}a, no-slip boundary conditions are applied to the AUV and propeller surfaces to make the velocity components and the turbulence kinetic energy of the AUV consistent with the wall.  Despite the wall boundary, the other quantities are set as far-ﬁeld quantities. The mesh of the total domain is shown in fig. \ref{fig:2k}b. The meshing for the propeller surrounding normal to the propeller axis is separately shown in fig. \ref{fig:2k}c. The mesh is made up of tetrahedral grid throughout the domain. 

\subsection{Grid independence study}
The standard wall function is used to model the ﬂow near the AUV and the propeller. The  logarithmic relation of the near-wall velocity is used here to determine the first layer thickness. The  logarithmic  law  for  the  mean velocity is valid when the distance from the ﬁrst point to the wall satisﬁes $30<y^{+}<300$.  The y+ value was derived as 70 for AUV and 124 for propeller, after three different simulation to optimize the y+ value, both of which falls in the reasonable regime of wall function approach. Therefore, the first layer thickness were calculated as 0.006 mm which is less than the maximum boundary layer thickness. A mesh independence test for the case having distance between the propeller and AUV of 0.08m and propeller speed of 1200 RPM was done. It has been shown in the Table 1. Different mesh size were taken as very coarse, coarse, fine and very fine. Even after the increase of number of grid beyond fine case, there is no significant change in the drag coefficient. From the analysis, grid size of 0.007 m (fine) is chosen for all other simulation cases. The number of grids were varied from 1.5-3.1 million. 
\begin{table}
\begin{center}
 \begin{tabular}
 {c  c c c c} 
 \hline
 \begin{tabularx}{\textwidth}{c*{5}{>{\raggedright\arraybackslash}X}}
\toprule
 Mesh resolution & Mesh size at the hull
and propeller surface 
 & Mesh size far from the hull and propeller & Number of cells & Hull total drag coefficient \\ [0.5ex] 
 \hline\hline
 Coarse & $0.010$  & $0.14$ & 1,550,336 & 0.071 \\ 
 \hline
 Mid & $0.008$ &  $0.14$ & 1,984,284 & 0.077  \\
 \hline
 Fine & $0.007$ &  $0.14$ & 2,305,374 & 0.085  \\
 
 \hline
 Very ﬁne & $0.006$ & $0.14$ & 3,129,532 & 0.087 \\ [1ex] 
\hline
\bottomrule
\end{tabularx}
\end{tabular}
\end{center}
\caption{Meshing strategy and mesh sensitivity study, the mesh size is in meter.\label{t1}}
\end{table}

\subsection{Solution Methodology and boundary conditions}
Moving reference frame were used for the numerical simulations. The domain was divided into two different parts- rotating region and static region. Rotating region was created around the propeller for providing required rotational velocity for the movement of the propeller and the other part is considered as the static region. The boundary conditions were taken as velocity inlet and pressure outlet for inlet and outlet respectively and the other walls were considered as no-slip boundaries. A ﬁnite-volume method is  applied  to discretize  the  governing equations. The hybrid and upwind scheme were used for the discretization of convective and diffusive terms. The implicit method was used for the transient part. All physical quantities are stored and computed at the cell centers. A pressure-based 3D unsteady solver was used in the numerical modeling.  The unsteady  formulation  is  1st-order  implicit.  The  SIMPLE algorithm is adopted to calculate the pressure ﬁeld. In addition, a standard  method  is  set for pressure discretization.  The  1st-order upwind  method  is  used  to  discretize  the  momentum,  the  turbulence kinetic energy and the turbulence dissipation rate. Twenty inner iterations have been done for ﬁxed per time step. Time step is taken as 0.001s considering the stability of solution from Courant number. Moreover,  the  minimum  residuals  are  maintained for  both  continuity  and  momentum  equations as $10^{-5}$ which match the usual convergence criteria.

\subsection{Governing equations}
Since we have used an open water tank in our experiments, the two phase flow mass and momentum conservation equations were used to model the motion of the AUV under free surface for replicating the actual flow phenomenon in the numerical simulations. We have the same equations as described in \cite{mansoorzadeh2014investigation}:
\begin{equation}
\begin{split}
\partial_{t} {(\alpha_i\rho_i)}+\nabla.(\alpha_i\rho_i U)=0, \quad i=1,2,
\end{split}
\end{equation}

\begin{equation}
\begin{split}
\alpha_i=\frac{V_i}{V}, \quad i=1,2,
\end{split}
\end{equation}

\begin{equation}
\begin{split}
\sum_{i} \alpha_i=1,
\end{split}
\end{equation}

\begin{equation}
\begin{split}
\sum_{i} \nabla.(\alpha_i U)=0.
\end{split}
\end{equation}

\begin{equation}
\begin{split}
\partial_{t} {(\rho_{m} U)}+\nabla.(\rho_{m} U*U)=\\\nabla.(-P+\mu_m((\nabla U)+(\nabla U)^T)), \quad i=1,2.
\end{split}
\end{equation}
Here, the velocity vector is denoted as $U$, the volume fraction of phase i is denoted as $\alpha_i$, $V_i$ is the volume of phase i. The density and viscosity are denoted as  $\rho_m$ and $\mu_m$ respectively, and the pressure acting on the flow is denoted as $P$. The discretization scheme for the convective term here was taken as first order upwind scheme. 

\subsection{Turbulence model}
The choice of turbulence model is an important factor for successful replication of the complex flow dynamics that arises because of stream line curvature or in the regions of high vorticity. In computational fluid dynamics simulations of turbulent flow, the central challenge is the prediction of the Reynolds stress tensor that arises after the ensemble averaging of the Navier-Stokes equations. Various researchers have developed several models to properly define the Reynolds stress. The basic CFD treatment of turbulence can be mainly categorized as eddy viscosity models\cite{pope2000,lumley1979}, Reynolds stress models\cite{panda2017,panda2018representation,mishra1,mishra2,panda2020areliabl}, Large eddy simulations\cite{pope2000} and direct numerical simulations\cite{pope2000}. Eddy viscosity models mainly encompasses two governing equations(one for the turbulence kinetic energy and another is the scale determining equation, usually the equation for dissipation rate as in the $k-\epsilon$ model). The Reynolds stresses are related to the instantaneous mean rate fo strain field by using the eddy viscosity hypothesis. Such eddy viscosity based turbulence models are exceedingly used in industrial applications, because of their simplicity and lower cost of application. But these models have led to unsatisfactory results in several circumstances, due to their use of the eddy viscosity hypothesis. The eddy viscosity hypothesis states that the Reynolds stresses are only related to the mean rate of strain, and ignore any dependence on the rate of rotation. It is a severe limitation in cases where there is significant streamline curvature or in presence of rotational flow field. Similarly, eddy viscosity based models assume that the Reynolds stress is dependent on the instantaneous rate of strain, thus ignoring any history (or memory) effects present in real life turbulent flows \citep{mishra2019estimating}, leading to unsatisfactory predictions. The other methods such as LES and DNS has higher cost of computation and very finer mesh is required for performing the numerical simulations, which is not possible with simple computer architecture. To overcome above difficulties we have performed numerical simulations with Reynolds stress models.           

Reynolds stress models solve transport equation for each and every component of Reynolds stress in the the flow field. We can study the detailed flow structure using these models. This eliminates the need of ad hoc definition of Reynolds stress (the turbulent viscosity hypothesis). The general form of the Reynolds Stress Transport Equation is given by \cite{panda2020review}
\begin{equation}
\begin{split}
&\partial_{t} \overline{u_iu_j}+U_k \frac{\partial \overline{u_iu_j}}{\partial x_k}=P_{ij}-\frac{\partial T_{ijk}}{\partial x_k}-\eta_{ij}+\phi_{ij},\\
&\mbox{where},\\ 
& P_{ij}=-\overline{u_ku_j}\frac{\partial U_i}{\partial x_k}-\overline{u_iu_k}\frac{\partial U_j}{\partial x_k},\\
&\ T_{kij}=\overline{u_iu_ju_k}-\nu \frac{\partial \overline{u_iu_j}}{\partial{x_k}}+\delta_{jk}\overline{ u_i \frac{p}{\rho}}+\delta_{ik}\overline{ u_j \frac{p}{\rho}},\\
&\eta_{ij}=-2\nu\overline{\frac{\partial u_i}{\partial x_k}\frac{\partial u_j}{\partial x_k}}  \\
&\phi_{ij}= \overline{\frac{p}{\rho}(\frac{\partial u_i}{\partial x_j}+\frac{\partial u_j}{\partial x_i})}\\
\end{split}
\end{equation}

In the above equation, $P_{ij}$ is the production of turbulence, $\eta_{ij}$ represents the dissipation process of turbulence, $T_{ijk}$ is the transport term and $\phi_{ij}$ is the pressure strain correlation term, which can be decomposed as slow and rapid pressure strain correlation. The slow term represent the turbulence/turbulence interactions and the rapid term represent the turbulence/mean strain interaction \citep{mishra3}. 

The pressure strain correlation term represents the non-local interactions and physics that is one of the key challenges of turbulence modeling. For instance, By proper definition of the pressure strain correlation term, accurate flow physics can be replicated in the numerical simulations. Unsatisfactory modeling of this term can lead to fundamental errors in simulations. For example it has been shown that dependent on the nature of the rotating flow field and it's alignment with the applied shear field, the pressure strain correlation term can have diametric effects on the turbulent flow \citep{mishra2019linear, mishra2015hydrodynamic}. The pressure strain correlation is also the term responsible for important secondary flow instabilities like the elliptic instability \citep{cambon1997energy, mishra6, godeferd1994detailed}. Accurate replication of such physics is essential for high fidelity predictions. 

As stated, the pressure strain correlation term can be decomposed into the slow (or nonlinear) and the rapid (or linear) pressure strain correlation. By precedent these 2 components are modeled separately. The most general form of the slow pressure strain correlation modeling expression has the form \citep{ssmodel}:
\begin{equation} 
\phi^{(S)}_{ij}=\beta_1 b_{ij} + \beta_2 (b_{ik}b_{kj}- \frac{1}{3}II_b \delta_{ij})
\end{equation}
The closure coefficients $\beta_1$ and $\beta_2$ can be the functions of second and third invariants of Reynolds stress anisotropy or can be a function of turbulent Reynolds number. $b_{ij}=\frac{\overline{u_iu_j}}{2k}-\frac{\delta_{ij}}{3}$ is the Reynolds stress anisotropy tensor, $II_b$ and $III_b$ are the second and third invariant of the Reynolds stress anisotropy respectively.

The general modeling expression for the rapid pressure strain correlation has the form \citep{mishra6}:
\begin{equation}
\begin{split}
& \phi_{ij}^{(R)}=S_{pq}[Q_1\delta_{ip}\delta_{jq}\\ &+Q_2(b_{ip}\delta{jq}+b_{jp}\delta{iq-2/3b_{pq}}\delta{ij})\\ &+Q_3b_{pq}b_{ij}+Q_4(b_{iq}b_{jp}-1/3b_{pk}b_{kq}\delta_{ij})
+\\ & Q_5b_{pl}b_{lq}b_{ij}+(Q_5b_{pq}+Q_6b_{pl}b_{lq})\\ &(b_{ik}b_{kj}-1/3II_{b}\delta_{ij}]
+\\ & \Omega_{pq}[Q_7(b_{ip}\delta{jq}+b_{jp}\delta_{iq})+Q_8b_{pk}(b_{jk}\delta{iq}+b_{ik}\delta{jq}+\\ & Q_9b_{pk}(b_{jk}b_{ik}+b_{ik}b_{jq})]
\end{split}
\end{equation}
where, $S_{ij}$ is the mean rate of strain, $W_{ij}$ is the mean rate of rotation and $K$ is the turbulent kinetic energy. $II_b = b_{ij}b_{ji}$ is the second invariant of the Reynolds stress anisotropy tensor. We have used a Low Reynolds number version of the linear pressure strain correlation model in our numerical simulations \citep{cd2017star,gibson1978ground}. This is often referred to as the two layer pressure strain correlation model. This model extends the linear model of \citep{gibson1978ground} so that it can be applicable to the near-wall sub-layer where viscous effects are dominant. More information on this pressure strain correlation model and its formulation is available in \cite{o2018assessment}. 

\section{Validation of the numerical model predictions with experimental results}
The experimental results of AUV with propeller for 800 RPM case is considered for validation. Reynolds
stress model with linear pressure strain correlation is considered for our simulations. The turbulence pa-
rameters were calculated from the measured velocity statistics. Mainly turbulence kinetic energy and two
components of Reynolds stresses were considered for validation. We have considered the shear stresses for
validations, since those have direct relationship with the skin friction evolution and subsequently on other
wall related parameters\citep{fukagata2002contribution, monte2011analysis}. As reported in literature the error asso-
ciated with the experimental results are with in 1 percent of the measured value\citep{voulgaris1998evaluation}. In figure 4a, the variation of turbulence kinetic energy is shown. The dotted and solid line correspond to only AUV and AUV with rotational 
flow field cases respectively. Since propeller imparts disturbances in the upstream  flow field of the AUV, a sharp increase in turbulence kinetic energy is observed for the case
of AUV with rotational flow field. A similar trend is observed for the Reynolds stresses. As reported in \cite{fukagata2002contribution}, the increased Reynolds stresses will have significant effect on the
evolution hydrodynamics parameters, those will be discussed in detail, in the subsequent sections.

\label{S:4}
\begin{figure}
\centering
\subfloat[Evolution of turbulence kinetic energy]{\includegraphics[height=7.2cm]{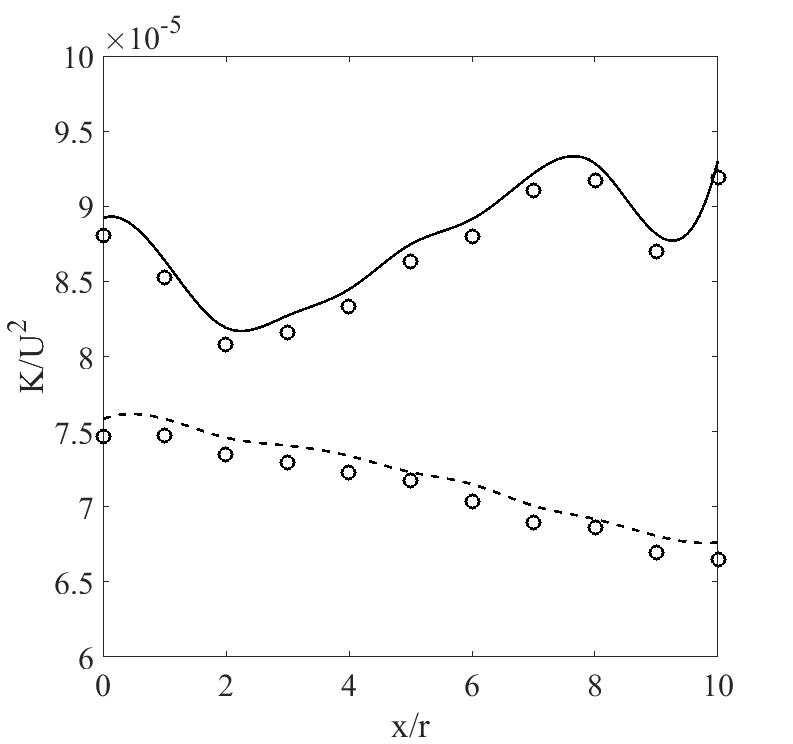}}\\
\subfloat[Evolution of Reynolds shear stress($R_{12}=\overline{uv}$)]{\includegraphics[height=7.2cm]{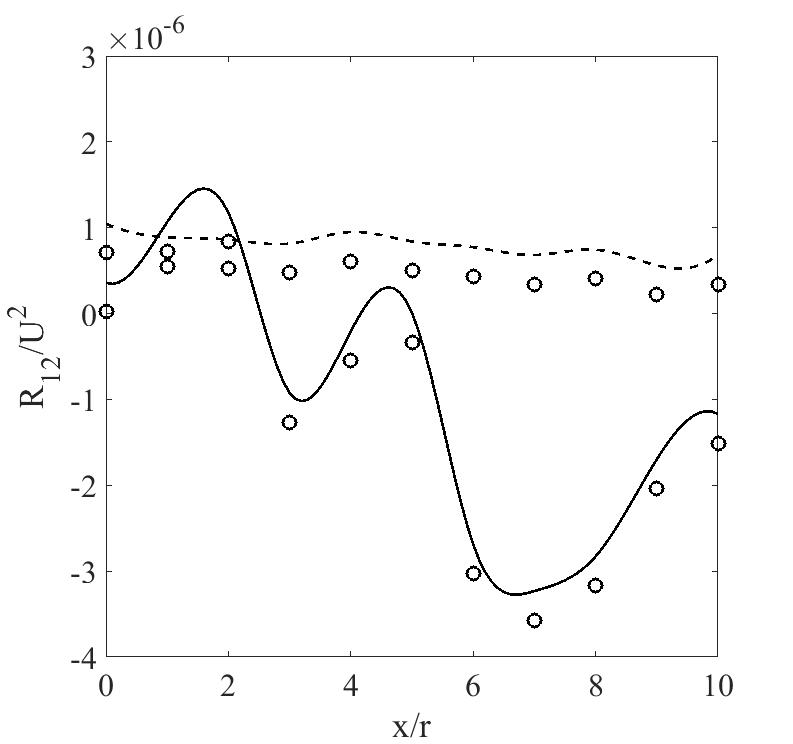}}\\
\subfloat[Evolution of Reynolds shear stress($R_{13}=\overline{uw}$) ]{\includegraphics[height=7.2cm]{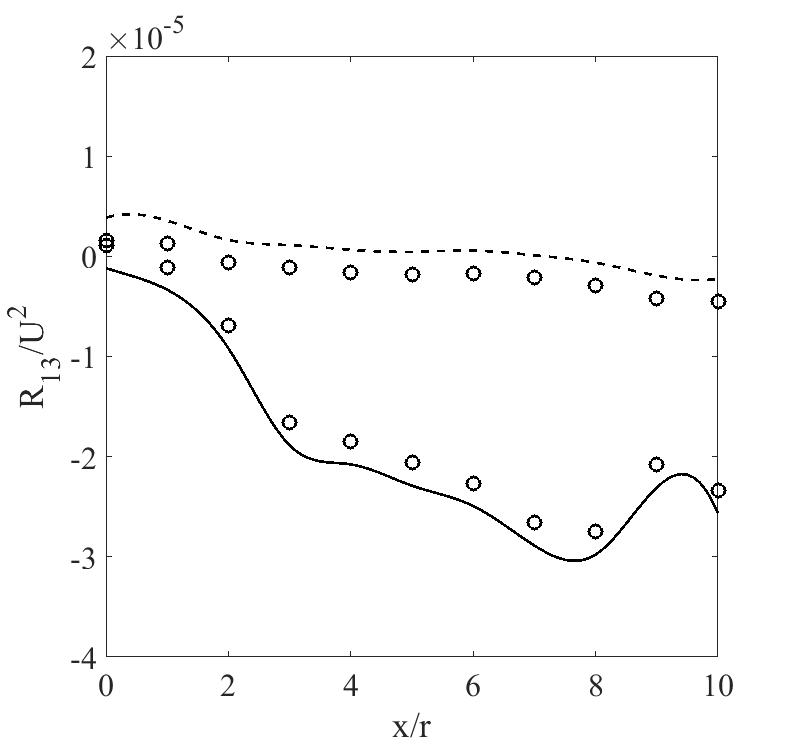}}\\
\caption{Comparison of Reynolds stress model predictions with the turbulence kinetic energy and  the components of Reynolds shear stress. In all the sub-figures, solid and dashed line represent evolution of the corresponding  parameter in presence and absence of rotating flow field respectively. The corresponding circles represent the experimental results \label{fig:3}}
\end{figure}

\section{Results and discussion}
\label{S:5}
In this section we will present numerical results of evolution of all the hydrodynamic parameters such as drag, pressure and skin friction coefficients with variation in the rotational velocity of the propeller and the distance of the propeller from the AUV. In the first and second subsection, the variation with rotational velocity of the propeller (which we have mentioned as strength of rotating field) and distance from the AUV hull will be presented respectively. The variations of hydrodynamic parameters for different angles of attack($\alpha$) and drift angles($\beta$) in the rotating flow field are also presented. 
\subsection{Effect of rotational field strength on the hydrodynamic parameters of the AUV}

\begin{figure}
\centering
\includegraphics[height=9.75cm]{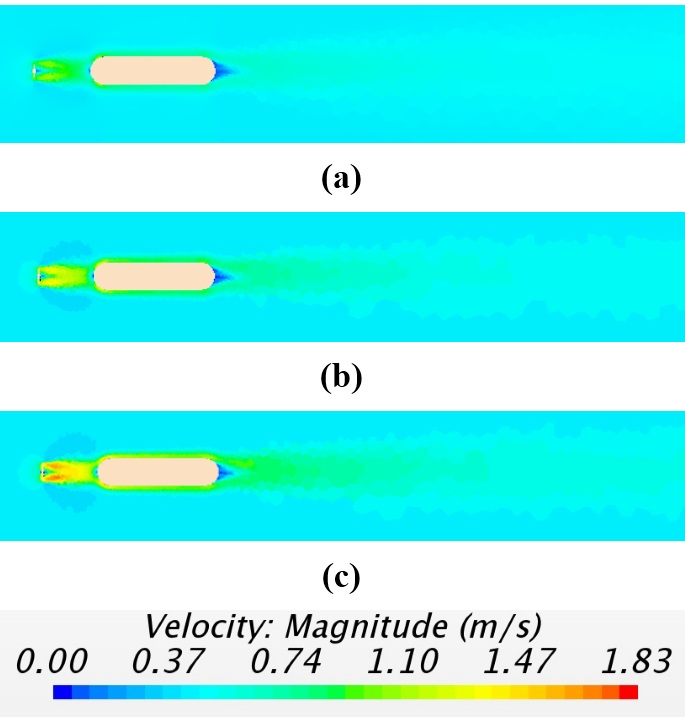}
\caption{Contours of velocity for different RPM of the propeller. a) 800, b) 1000 and c) 1200 RPM respectively.} \label{fig:4}
\end{figure}

Figure \ref{fig:4} presents contours of the velocity magnitude across the propeller and AUV for different ranges of RPM of the propeller. For all the three figures the fluid flow velocity was maintained as 0.42 m/s. It is clearly visible from the contour diagram that, with increase in rotational speed of the propeller, the strength of re-circulation zone past the AUV increases. It is well described in the literature, that the size and strength of re-circulation zone plays important role in the evolution of hydrodynamic parameters across the structure considered \citep{son2010effect}. Prior research has established that the strength of the recirculation zone has direct effect on the pressure drag coefficient (associated with form drag) for AUV designs \citep{de2014numerical}. Thus, higher strength of the ambient rotating flow field leads to higher form drag experienced by the AUV. 

\begin{figure}
\centering
\includegraphics[height=8cm]{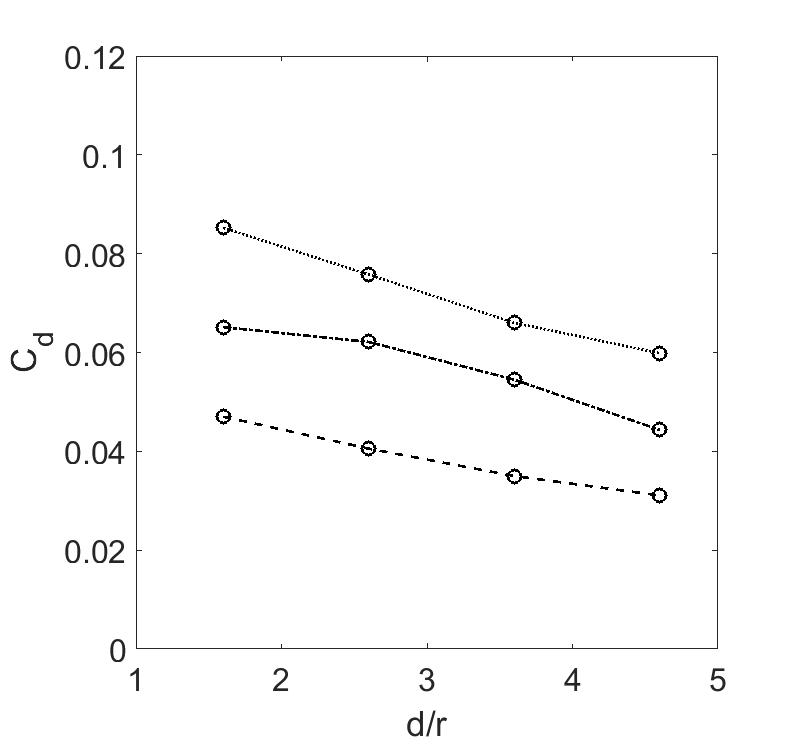}
\caption{Variation of drag coefficient with distance from propeller, Dashed, dashed-dot and dotted lines represent 800, 1000, 1200 RPM respectively.} \label{fig:5}
\end{figure}

Figure \ref{fig:5} represent the interesting findings on the drag evolution with propeller RPM and distance (distance of propeller from AUV) variations. The dashed, dashed-dot and dotted lines represent variation of drag coefficient for 800, 1000, 1200 RPM respectively. The radius of the AUV hull $r$ was used to non-dimensionalize all the parameters in the x-axis and $d$ is the distance of AUV from propeller. It is clear from the figure \ref{fig:5} that, with increase in rotational speed of the propeller, there is a sharp increase in drag of the AUV (From a separate numerical simulation for AUV in absence of propeller the drag coefficient was appeared to be 0.02, which is much smaller than the drag of the AUV in presence of the propeller). The drag was enhanced because of the increase of the strength of vortex shedding (The strength of vortex shedding has a direct effect on the evolution of drag along a body \citep{son2010effect}). From velocity contours in figure \ref{fig:4} it can be observed that the rotational speed of the propeller is enhancing the strength and size of recirculating zone in wake region of the AUV hull, this because of increase in strength of vortex shedding \citep{bakic2003experimental,son2010effect}.

\begin{figure}
\centering
\includegraphics[height=8cm]{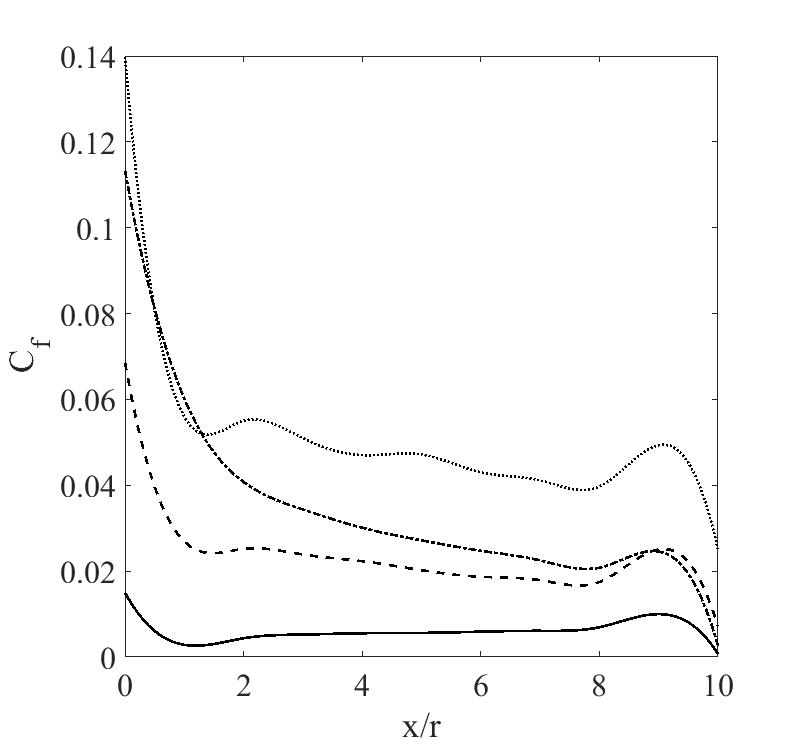}
\caption{Variation of skin friction coefficient along the AUV for different RPM of the propeller. Dashed, dashed-dot and dotted lines represent 800, 1000, 1200 RPM respectively. The solid line represent skin friction coefficient of AUV in absence of propeller. The propeller distance from the AUV is 0.13 meter.} \label{fig:6}
\end{figure}

\begin{figure}
\centering
\includegraphics[height=8cm]{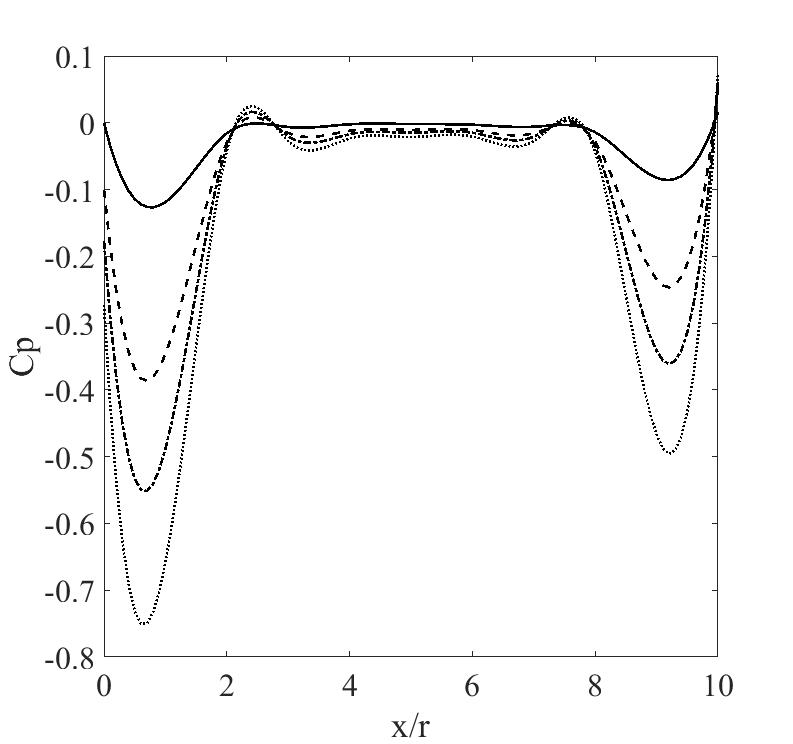}
\caption{Variation of pressure coefficient along the AUV for different RPM of the propeller. Dashed, dashed-dot and dotted lines represent 800, 1000, 1200 RPM respectively. The propeller distance from the AUV is 0.13 meter.} \label{fig:7}
\end{figure}

In figure \ref{fig:6} and \ref{fig:7} the variation skin friction and pressure coefficients are presented. In both the figures the solid lines represent corresponding variation of the parameter along the AUV hull in absence of propeller (in absence of rotational flow field). The dashed lines, dashed-dot line and dotted lines represent the variations at 800, 1000 and 1200 RPM of the propeller respectively. It can be seen from the figures, there is sharp increase of both of the coefficients with increase in rotational speed of the ARM. In both of the figures, x/r=0 and 10 corresponds to two extreme ends of the AUV. x/r=0, is at the propeller side of the AUV. The magnitudes of the skin friction and pressure coefficients are larger at the propeller side. As the rotational strength is decreasing towards the end of the propeller, a decrease in both of the coefficients is noticed.

\subsection{Effect of propeller distance on the hydrodynamic parameters of the AUV}
\begin{figure}
\centering
\includegraphics[height=8cm]{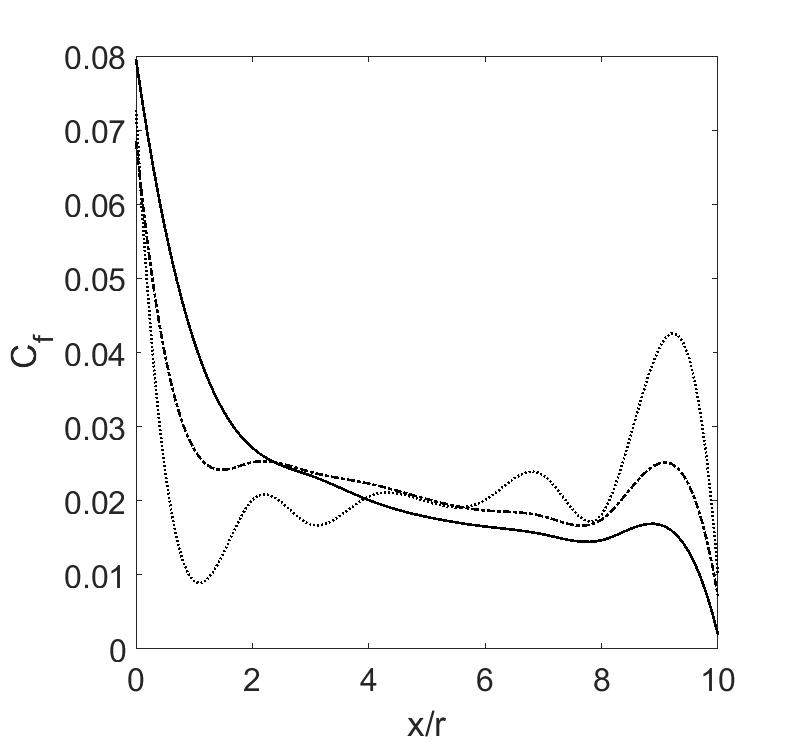}
\caption{Variation of skin friction coefficient along the AUV with variation of distance of the propeller. Solid, dashed-dot and dashed lines represent 0.08, 0.13, 0.23 meter respectively for 800 RPM of propeller.} \label{fig:8}
\end{figure}

\begin{figure}
\centering
\includegraphics[height=8cm]{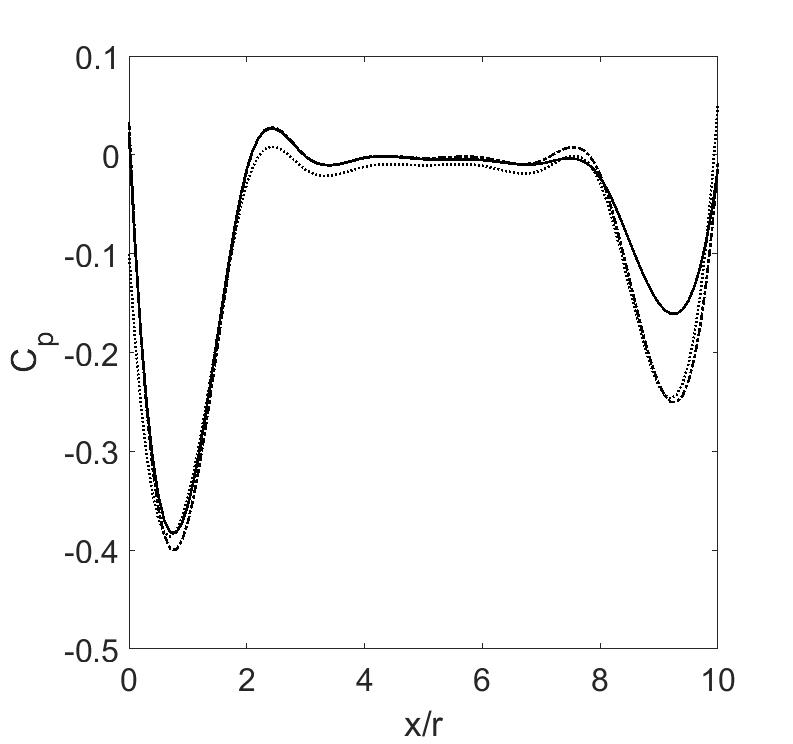}
\caption{Variation of pressure coefficient along the AUV with variation of distance of the propeller. Solid, dashed-dot and dashed lines represent 0.08, 0.13, 0.23 meter respectively for 800 RPM of propeller.} \label{fig:9}
\end{figure}

In this subsection, the variation of hydrodynamic parameters of the AUV will be presented by varying the distance of the propeller from the AUV hull. Figure \ref{fig:8} present the variation of skin friction along the AUV hull. The solid, dashed-dot and dashed lines represent the distances 0.08, 0.13 and 0.23 meters respectively. We could not conduct the experiments for 0.18m case, since there was a obstruction over the tank, for which, the measuring instrument could not be fixed at the appropriate location. It can be seen from figure \ref{fig:8} that when the propeller is near the AUV, the skin frcition is more. A gradual decrease in skin friction is observed with increase in distance of propeller from AUV. The same trend is also observed for the pressure coefficient evolution in figure \ref{fig:9}.

\subsection{The effects of angle of attack on the hydrodynamic parameters of the AUV in presence of rotating flow field}
\begin{figure}
\centering
\includegraphics[height=8cm]{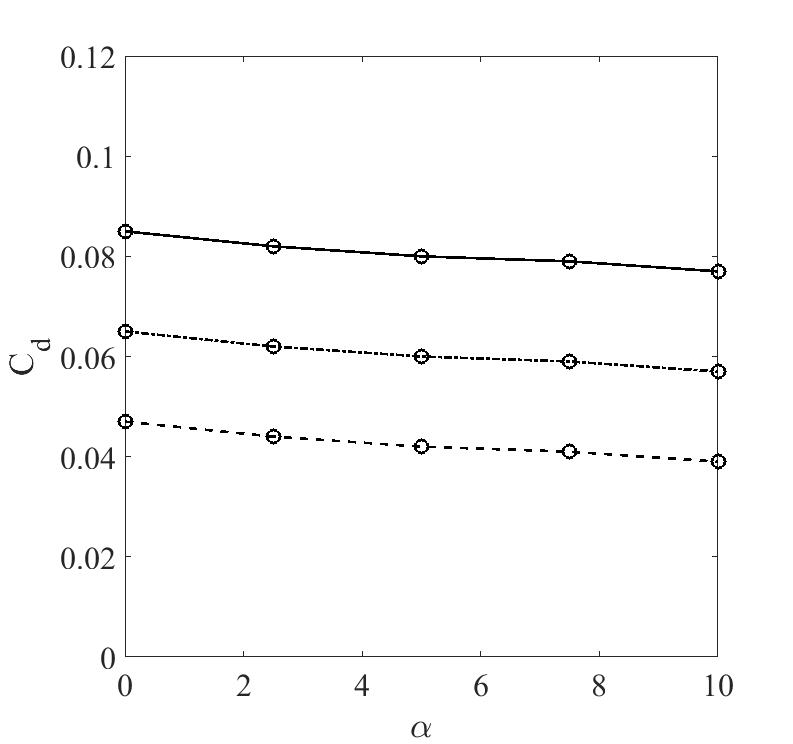}
\caption{Variation of drag coefficient of AUV with variation of angle of attack for different RPM of the propeller. Dashed, dashed dot and solid lines represent 800, 1000 and 1200 RPM respectively.} \label{fig:10}
\end{figure}

\begin{figure}
\centering
\includegraphics[height=8cm]{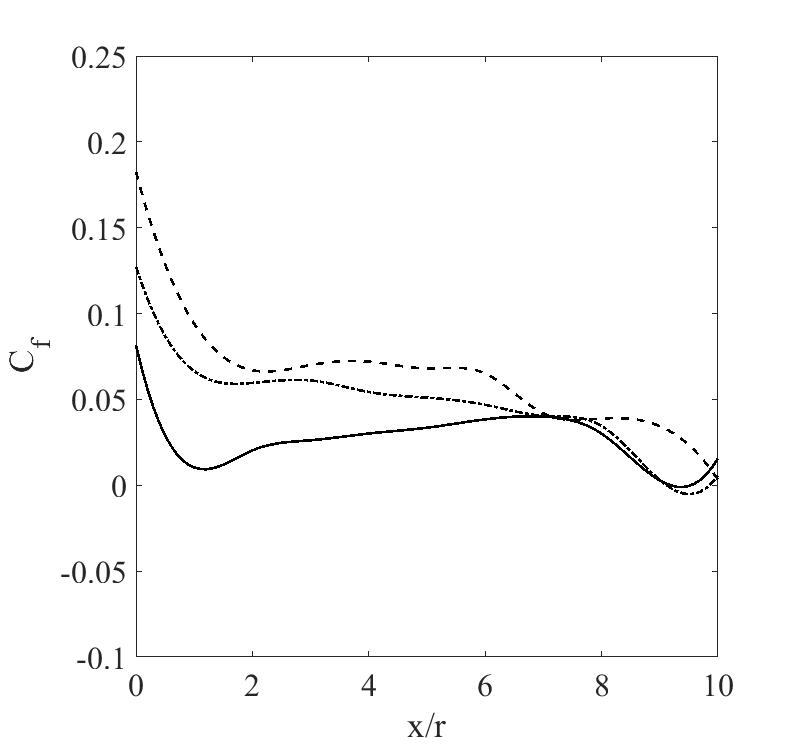}
\caption{Variation of skin friction coefficient along the AUV for different angles of attack. Dashed, dashed-dot and dotted lines represent 0, 5, 10 degrees respectively. The RPM of the propeller was 1200 and distance from the AUV hull was 0.08 meter.} \label{fig:11}
\end{figure}

\begin{figure}
\centering
\includegraphics[height=8cm]{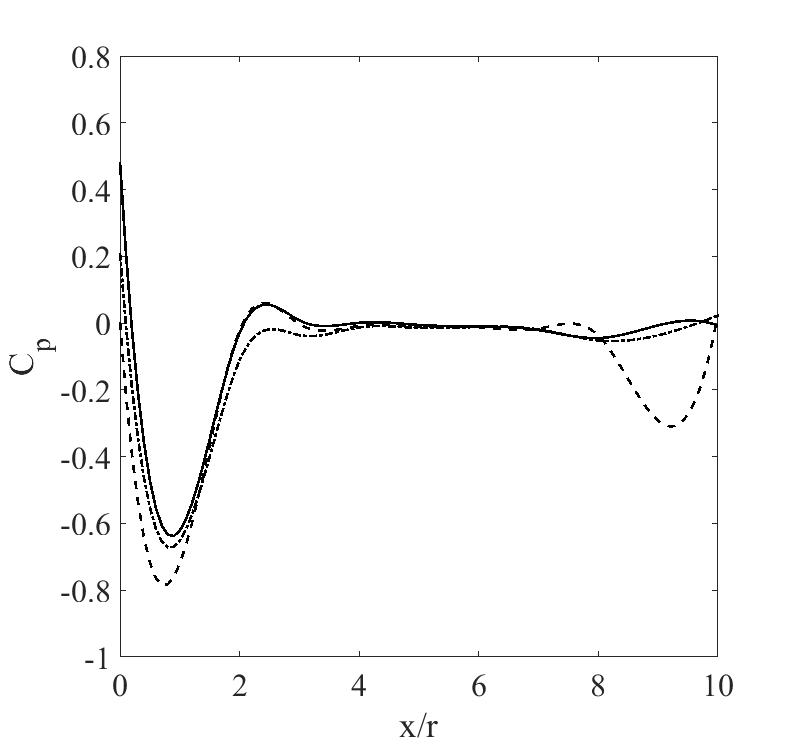}
\caption{Variation of pressure coefficient along the AUV for different angles of attack. Dashed, dashed-dot and dotted lines represent 0, 5, 10 degrees respectively. The RPM of the propeller was 1200 and distance from the AUV hull was 0.08 meter.} \label{fig:12}
\end{figure}

In Figure \ref{fig:10}, the variation of drag coefficients with different angles of attack are presented. The results were also contrasted against results of RPM variations of the propeller. The dashed, dashed-dot and solid lines present drag coefficient evolution for 800, 1000 and 1200 RPM of the propeller respectively. It is clear from the figure that, the drag of the AUV hull decreases with increase in angle of attack in presence of the rotating flow field. The slope of the drag evolution is very small. The skin friction and pressure   coefficient evolution for different angles of attack are presented in figures \ref{fig:11} and \ref{fig:12}. Although we have performed several simulations with different RPM and distance of propeller from AUV hull, here we have presented $C_p$ and $C_f$ evolution for the case of 1200 RPM and 0.08 meter distance of propeller. In presence of rotational flow field the skin friction coefficient of the AUV was reduced with increase in angle of attack. The angle of attack variation has minimal effect on the pressure coefficient evolution. Usually the drag of AUV increases with increase in angle of attack (Jagadeesh and Murali\cite{jagadeesh2009experimental}). The reverse trend of drag evolution is due to the reduction of friction drag of the AUV hull. As noticed from the figure \ref{fig:10} the drag reduction is very minimal with increase of angle of attack.     


\subsection{The effects of drift angle on the hydrodynamic parameters of the AUV in presence of rotating flow field}
\begin{figure}
\centering
\includegraphics[height=8cm]{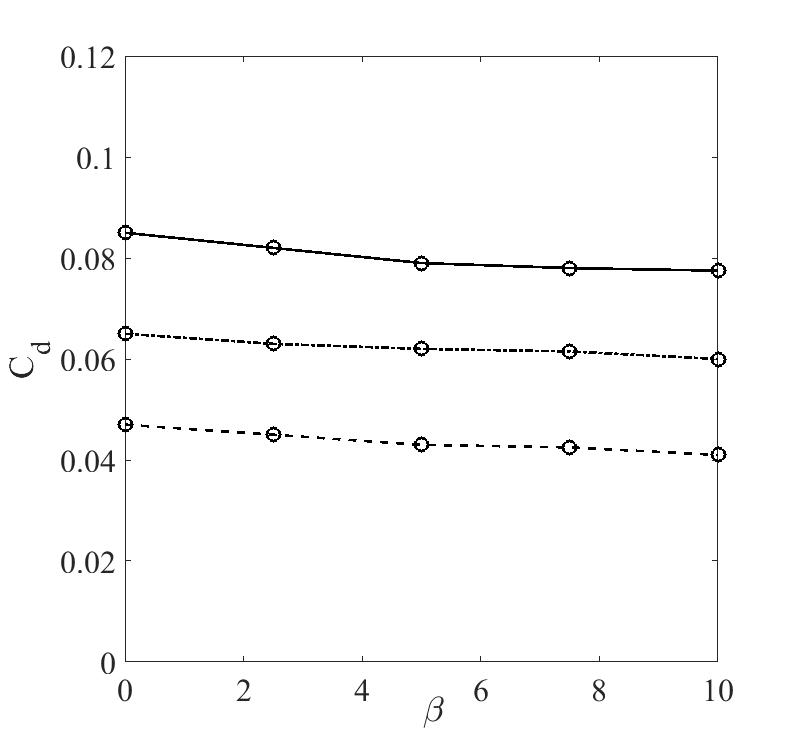}
\caption{Variation of drag coefficient of AUV with variation of drift angle for different RPM of the propeller. Dashed, dashed dot and solid lines represent 800, 1000 and 1200 RPM respectively.} \label{fig:13}
\end{figure}

\begin{figure}
\centering
\includegraphics[height=8cm]{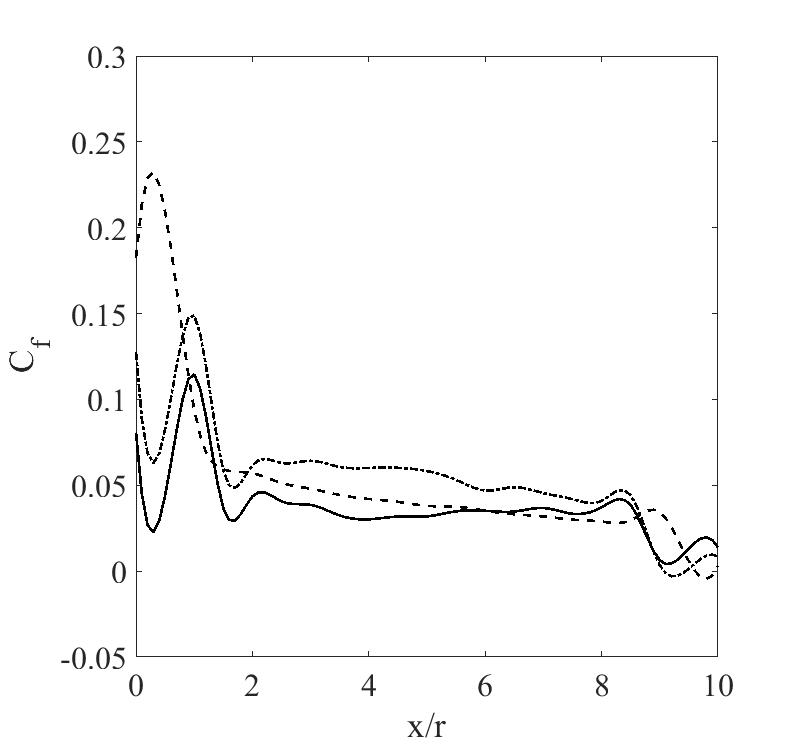}
\caption{Variation of skin friction coefficient along the AUV for different drift angles. Dashed, dashed-dot and dotted lines represent 0, 5, 10 degrees respectively. The RPM of the propeller was 1200 and distance from the AUV hull was 0.08 meter.} \label{fig:14}
\end{figure}

\begin{figure}
\centering
\includegraphics[height=8cm]{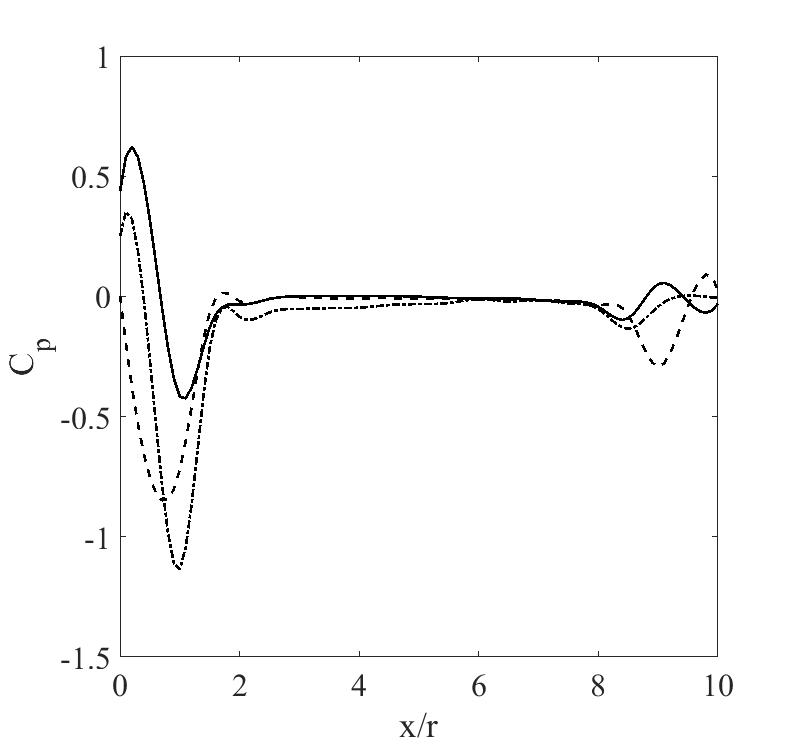}
\caption{Variation of pressure coefficient along the AUV for different drift angles. Dashed, dashed-dot and dotted lines represent 0, 5, 10 degrees respectively. The RPM of the propeller was 1200 and distance from the AUV hull was 0.08 meter.} \label{fig:15}
\end{figure}

The evolution of drag for different drift angle with the variation of RPM of the propeller is presented in figure \ref{fig:13}. Basically a "drift angle" is developed when a underwater vehicle takes a turn. The drift angle at a point along the length of the AUV is defined as the angle between initial central line of the AUV the final central line after a turning\citep{bridgesdh2003experimentalinvestigationoftheflowpastasubmarineat}. The drag coefficient of the AUV hull was decreased With increase in drift angle. For all three RPM of the propeller, similar trend was observed for the evolution of drag. This is also because of the reduction of skin friction drag of the AUV. In Figures \ref{fig:14} and \ref{fig:15} the evolution of skin friction and pressure coefficient of the AUV are presented for drift angle variations.         


\newpage
\section{Concluding remarks}
In this article, we provide a detailed study on the effect of the rotational flow field of varying strengths on the hydrodynamic characteristics of an AUV. The rotating flow field was produced by placing a propeller in the flow field of the AUV in a water tank. The experimental results of turbulence statistics in the vicinity of the AUV were used to validate the numerical model predictions and finally the variation of drag, pressure and skin friction coefficient evolution were analyzed along the AUV for different rotational flow fields. The effect of distance of propeller on the hydrodynamic coefficients of the AUV were also studied. It was noticed that in presence of rotational flow field the drag of the AUV increases. A sharp increase of drag of the AUV was observed with increase in rotational strength and decrease of distance of the propeller from the AUV. The angle of attack and the drift angle of the AUV were also varied to study the variations of the hydrodynamic parameters. The drag coefficient of the AUV was reduced both with increase in angle of attack and drift angle. This is mainly because of the reduction of skin friction drag of AUV in the rotating flow field. The experimental and numerical results presented in this article, will be useful for design optimization of AUVs operating in deeper oceans and in the regions of high vorticity.




\newpage
\bibliographystyle{elsarticle-harv}\biboptions{authoryear}
\bibliography{asme2e.bib}







\end{document}